# Microstructural evolution of a low-alloy steel / nickel superalloy dissimilar metal weld during post-weld heat treatment

C.V. da Silva Lima[a], M. Verdier[a], F. Robaut[a], J. Ghanbaja[b], G. Badinier[c], T. Marlaud[c], C. Tassin[a], H.P. Van Landeghem[a,1]

[a] Univ. Grenoble Alpes, CNRS, Grenoble INP, SIMAP, F-38000 Grenoble, France

[b] Institut Jean Lamour UMR 7198 CNRS, Université de Lorraine, BP 70239, F-54506, Vandœuvre les Nancy, France

[c] Framatome Lyon, 10 rue Juliette Récamier, F-69456 Lyon Cedex 06, France

Abstract: The microstructural evolution of a dissimilar metal weld (DMW) obtained by narrow-gap gas tungsten arc welding (NG-GTAW) was investigated after it was subjected to a post-weld heat treatment (PWHT). The case studied here is a joint between low-alloy steel pipes and a stainless steel steam generator using a nickel based alloy as filler material. The fusion boundary that was the focus of this work was that between the low-alloy steel (2.25Cr-1Mo) and the nickel alloy (alloy 82). The difference in matrix phase and chemical composition between the two alloys leads to a large difference in chemical potential for carbon, which is mobile at the PWHT temperature. A number of advanced characterization techniques were used to assess the gradient of composition, hardness and microstructures across the fusion line, both as welded and after PWHT. This complete analysis permits to highlight and understand the main microstructural changes occurring during the PWHT.

## I. Introduction

Dissimilar metal welds (DMWs) are a common occurrence in power plants where the need to join different alloys or alloy grades is frequent due to elaborate piping design (Brentrup and Dupont, 2013). Meeting stringent safety requirements over the long life span of the facilities while minimizing capital expenditure requires using optimal alloys for the target components. The case studied here is that of a narrow-gap (NG) DMW prototype that would join the steam generator hot collector to the steam circuit in a fourth generation, sodium-cooled nuclear power plant. The pipes of this circuit are planned to be manufactured out of 11CrMo9-10, a low alloy steel (LAS) with excellent creep resistance. Such components will be arc welded to the steam generator made of austenitic alloy 800H (UNS N08810) using nickel-based alloy 82 (NiCr20Mn3Nb) as filler material. The resulting junction belongs to the broader class of ferritic/austenitic DMWs. The fusion boundary (FB) of particular interest here is the one between the LAS (ferritic) and the filler metal (austenitic), where the large mismatch in properties plays a critical part in the performance of the weld (Nivas et al., 2017).

Alloy 82 has been used in earlier-generation nuclear power plants to join SA508 Cl.3 pressure vessel (≈ 18MND5) with the 316L primary circuit before being progressively phased out and superseded by alloy 52 (Hänninen et al., 2006), which displays an enhanced resistance to primary water stress corrosion cracking due to its higher chromium content (P.Andresen, 2004). However, the steam environment of the present case does not present the risk of stress corrosion cracking observed on DMWs with alloy 82 in PWR primary water. Moreover, this same higher chromium content leads to a more pronounced transfer of carbon from the LAS side to the weld side. This phenomenon, which is known to play a major part in DMW failures (DuPont, 2010), is exacerbated at the higher operation temperatures (≈ 530°C) of sodium cooled reactors. Finally, long term exposure of filler metal alloy 52, whose chemical composition is adapted from bulk metal alloy 690, to these higher operation temperatures results in ageing embrittlement due to short-range ordering (SRO) and/or long-range ordering (LRO) (Martinsson, 2006). This atomic ordering transformation is based on $Ni_2Cr$ super-lattice (Marucco, 1994). Its degree of order and kinetics depend on chemical composition of nickel-base superalloy(Marucco, 1995). Due to its lower-chromium and higher-iron composition, alloy 82 is less sensitive to this ordering ageing embrittlement.

---
[1] Corresponding author. E-mail: hugo.van-landeghem@grenoble-inp.fr

NG gas tungsten arc welding (GTAW) is a joining process used notably for DMW that renders superfluous the use of buttering layers on the LAS nozzle. It presents a number of advantages over previous processes in terms of both cost and performance. However, it has been shown that this change in weld joint design can lead to a different response to PWHT (Nevasmaa et al., 2013). For the widespread combination of SA508 and 316L joined using alloy 52, PWHT following NG-GTAW does not improve (Ahonen et al., 2016) and can even worsen (Nevasmaa et al., 2013) the fracture resistance of the zone near the FB between SA508 and alloy 52. This finding highlights the importance of understanding the effects of PWHT on the microstructure and its associated properties.

The SA508/alloy 52 combination is widespread in nuclear power plant applications and these ferritic/austenitic DMWs have been thoroughly studied, especially their mechanical behavior (Keim et al., 2015) in the as welded and PWHT conditions. However, to the knowledge of the present authors, the amount of published work regarding joints of 2.25Cr-1Mo steel with nickel-based alloy 82 is less definitive regarding the effect of PWHT on the microstructure. In addition, PWHT vary in both temperature and duration from one study to another, from 700°C for 1 h(Laha et al., 2001) to 3 h(Parker and Stratford, 2000), to 630°C for 7 h (Elrefaey et al., 2018). The purpose of the present paper is thus to investigate the microstructural changes induced by PWHT and their consequences on mechanical properties of such joints. It is known that in such joints, the key part is the direct vicinity of the ferritic/austenitic FB, within a few tens of microns (Bhaduri et al., 1995). A thorough characterization, involving multiple techniques, of this region in as-welded and PWHT samples is presented in the following to highlight the evolution of the microstructure, the composition gradients and the resulting mechanical properties. The results are then discussed in terms of the chosen PWHT conditions to assess whether they could be altered to yield a more optimal microstructure of the near-FB zone.

## II. Experimental

### A. Base materials and processing

The materials involved in this study were a 11CrMo9-10 LAS forged-plate, an alloy 800H hot-rolled plate and, as filler metal, a 1.2 mm diameter alloy 82 solid wire. Their compositions are given in Table 1.

A DMW mock-up was made of two plates of parent metal having the same dimensions: 500 × 150 × 25 mm³, assembled by a multi-pass butt joint. To limit deformations during welding, the assembly was clamped using three hangers. Welding was performed in flat position (1G) with a hot-wire GTAW process in narrow groove configuration (1 pass per layer) and using Arcal 37 (70% He 30% Ar) as shielding gas. The main welding parameters are:

- GTAW hot-wire – pulled passes,
- No preheating or post-weld "bake-out",
- maximum interpass temperature: 180°C,
- welding speed: 100 cm/min,
- filler feed rate: ≈ 140 cm/min,
- fixed welding voltage: 12V,
- direct and pulsed welding current.

The joint was then PWHT at 700°C for 2 h. The PWHT conditions were chosen in agreement with recommendations from guidelines for welding creep strength-enhanced ferritic alloys on heavy wall weldments, which are 2 h at ≈ 730°C (Coleman, 2007) and 2-4 h between 670°C and 720°C (Hilkes and Gross, 2013), respectively, for 11CrMo9-10 LAS.

After welding, the root passes and the finishing passes were removed by machining, in order to have a remaining thickness of 25mm welded with the same welding parameters, as illustrated by the metallography of the joint in Figure 1.

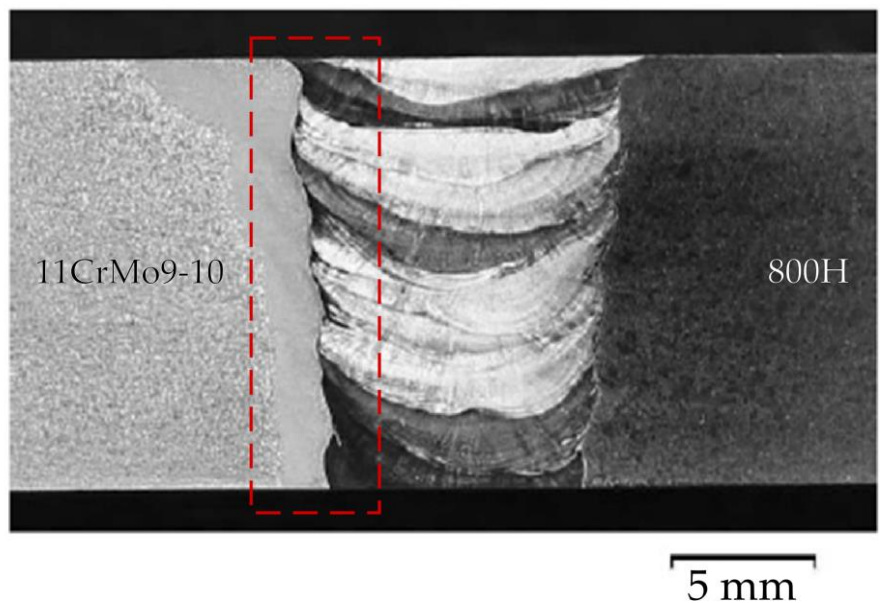

*Figure 1: Macrograph of the studied DMW between 10CrMo9-10 low-alloy steel and alloy 800H, joined using alloy 82 wire as filler metal. The austenite/ferrite fusion boundary is highlighted in the red dashed box.*

| Alloy | C | Cr | Ni | Mn | Si | Fe | Mo | S | P | Cu | Al | V | Co | Ti | Nb | Ta |
|---|---|---|---|---|---|---|---|---|---|---|---|---|---|---|---|---|
| 11CrMo9-10 | 0.12 | 2.18 | 0.15 | 0.45 | 0.23 | Bal. | 0.94 | 0.005 | 0.010 | 0.10 | 0.024 | 0.01 | N/A | N/A | N/A | N/A |
| 800H | 0.07 | 20.58 | 30.31 | 0.60 | 0.52 | Bal. | N/A | 0.002 | 0.011 | 0.07 | 0.25 | N/A | 0.11 | 0.33 | 0.01 | N/A |
| 82 | 0.01 | 20.6 | Bal. | 3.2 | 0.02 | 0.2 | N/A | 0.002 | 0.002 | 0.01 | N/A | N/A | 0.01 | 0.33 | 2.70 | 0.01 |

*Table 1: Composition of alloy 800H, 11CrMo9-10 low-alloy steel and alloy 82 used in this study.*

### B. Sample preparation

Samples prepared for metallographic observations were first cold mounted in an epoxy resin before being ground on SiC paper down to P2400 grit. They were then polished using diamond suspensions down to 1 μm. In order to reveal the microstructure of both 11CrMo9-10 LAS and alloy 82, the samples were dual etched. First, the LAS side was etched using Nital 3 % (3 % $HNO_3$ in methanol) for 20 to 90 s. Then, the weld side, unaffected by the previous step, was electro-etched in a chromic acid solution (10 % $Cr_2O_3$ in water) under a potential of 3 V, resulting in a current density of about 0.5 A.cm$^{-2}$ for 3 s. After etching, the samples were observed using an Olympus BX60M light microscope and a Zeiss Gemini SEM operated at 10 kV.

To ensure that the data from the various small-scale characterization techniques could be correlated in a straightforward manner, a number of region of interests (ROIs) were identified and marked on the etched samples. The marking was performed using micro-hardness indents with a 200 g load.

Given the surface requirements for EPMA and nano-indentation, the samples were appropriately ground again to remove all traces of etching before being polished using diamond suspensions down to 1 µm. Just before being loaded into the EPMA, the samples and the carbon analysis standards went through a final polishing step using OPA followed by thorough rinses with water then ethanol.

### C. Electron probe micro-analysis (EPMA)

The composition profiles were measured in the ROIs with a Cameca SX50 microprobe whose automation was performed with SAMx software. Carbon was analyzed with a tungsten/silicon multilayer monochromator (2d spacing = 9.5 nm -crystal PC2-) at an accelerating voltage of 13 kV with a 900 nA beam current for improved counting statistics. In this study, a liquid nitrogen cold trap and a low-pressure jet of oxygen onto the specimen were used to reduce carbon contamination. According to ISO standard 16592:2012, the carbon analysis procedure was based on the determination of a calibration curve for the carbon K$\alpha$ line intensity as a function of carbon content in the range of 0-1 wt.%. This regression curve is a straight line given by:

$$cts = A \times w_C + B$$

where $cts$ is the count number at the intensity peak maximum, $w_C$ is carbon weight content, and A and B are fit coefficients. In this procedure, background intensity is not measured. The set of homogeneous standards used to determine the calibration curve, consisted of pure iron and fully martensitic steels with, respectively, 0.2 wt.% C (20NiCrMo2), 0.42 wt.% C (42CrMo4), 0.98 wt.% C (100Cr6), and 0.99 wt.% C (100Cr6) (Robaut et al., 2006). Carbon content in the investigated samples was simply calculated from the regression curve coefficients A and B.

Cr, Mn and Ni contents were measured at 13 kV and 100 nA, with LiF monochromators (2d spacing = 0.4 nm). The classic quantitative analytical procedure was taken into account using the ratio of measured intensities of the K$\alpha$ line from the sample and the pure metallic standards. Background intensity was measured and deducted from the peak intensity.

### D. Nanoindentation

Local mechanical properties were measured using a commercial machine (MTS-XP) along two regularly spaced grids, as shown in Figure 2. Those two grids consisted of a coarse matrix (typically 30 × 2 indents at 450 nm depth, separated by 20 µm) along the EPMA profile and a fine matrix (typically 30 × 6 indents at 120 nm depth, separated by 2.5 µm in X and 5 µm in Y) straddling the FB in between the rows of the coarse matrix. Additionally, two reference matrices for both weld metal and LAS (3 × 2 indents, 450 nm depth) were recorded more than 1 mm away to obtain average values of the properties far from the FB. Tests were carried out using a Berkovich indenter to a predefined depth with a constant strain rate of $10^{-2}$ s$^{-1}$, with a continuous stiffness measurement (using a superimposed 2 nm displacement amplitude at 80 Hz). Standard analysis using Oliver and Pharr method are used to obtain both Young modulus (E) and hardness (H) at specified depth. The sample preparation used for EPMA provides the surface roughness required to conduct measurements at small depths (100 nm for the smallest one).

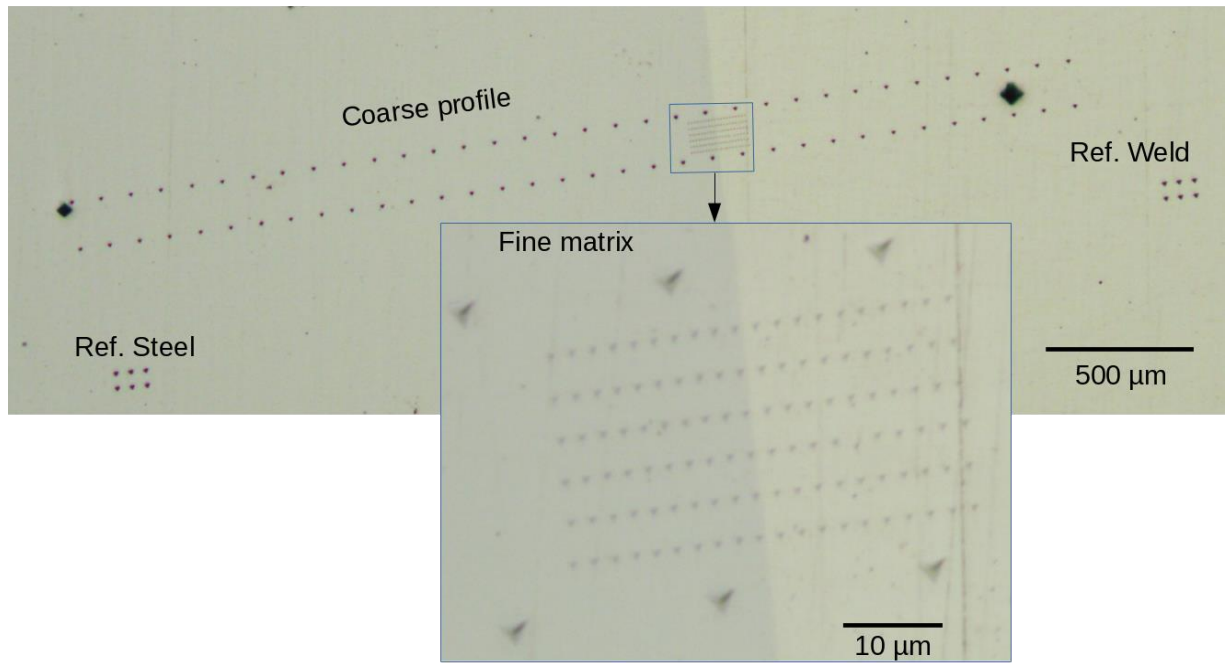

*Figure 2: Light micrographs showing the typical arrangement of nano-indents across the fusion boundary in two matrices, coarse and fine. The Vickers micro-indents used to mark the region of interest are also visible on both sides of the boundary.*

### E.     Transmission Electron Microscopy (TEM)

Lamellae for TEM observations were lifted out from the samples using an FEI Helios NanoLab 650 dual beam microscope. The lamellae were then further thinned in a Gatan PIPS II operated at 500 V, for a total of 2 min on each side, in 30 s steps. Before the observations, the lamellae were cleaned in a Gatan Solarus plasma cleaner for 4 min in an $Ar/H_2$ mixture followed by for 4 min in $O_2/H_2$. The specimens were observed in a JEOL 2100F equipped with a Centurio EDS detector and in a JEOL AccelARM both operated at 200 kV.

## III.    Results

### A.     Microstructural evolution

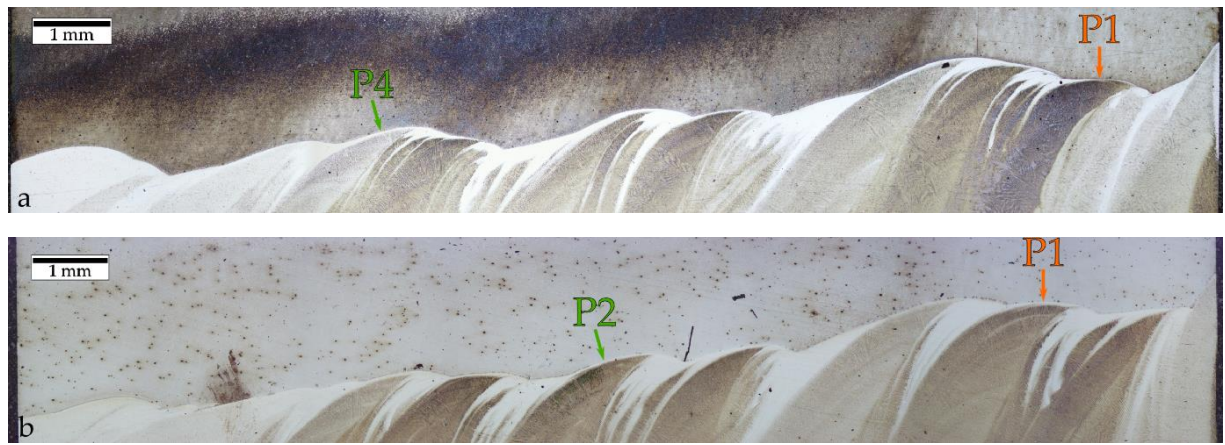

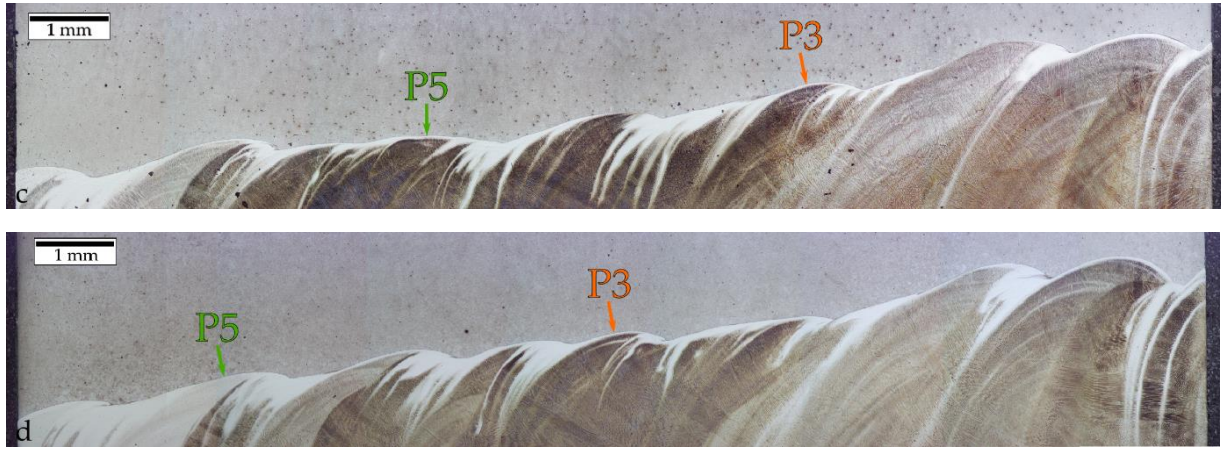

*Figure 3: Light micrographs of samples a) M1 and b) M2 in the as-welded condition, and samples c)M3.1 and d)M3.2 in the PWHT condition. Special positions used for finer characterizations of the near fusion boundary area are labeled here. Orange labels were used for regions with large white layers and green labels were used for small white layers. The weld passes shown in those micrographs were stacked from left to right.*

An overview of the microstructures of the welds before and after PWHT is presented in Figure 3. It can be seen that the welds present the usual aspect of NG GTAW weld where individual passes can be identified by the periodic change of the white layer (WL) directly adjacent to the FB from a thin parallel-sided layer to a feathery feature, reaching deep into the weld side. This layer, unaffected by electro-chromic etching, is a marker of the Fe-rich (> 20 wt%) region of the weld. Its shape and appearance reveal the local mixing of LAS and filler metal during welding, as shown in Figure 2. This general aspect is unaffected by the PWHT. In order to establish an appropriate comparison at a finer scale, ROIs in the as-welded and PWHT samples were selected where the WLs have comparable thicknesses.

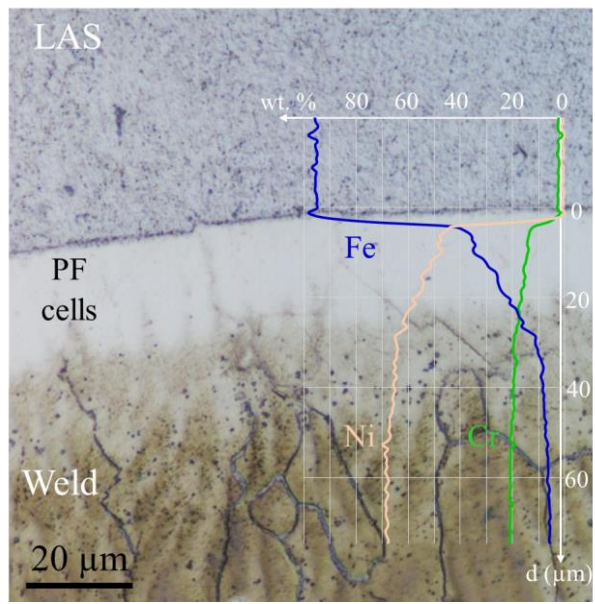

*Figure 4: Light micrograph of the fusion boundary (sample M3.2) in the PWHT condition with overlaid EDS profiles for Fe, Ni and Cr. Solidification microstructures in the weld are evidenced by electro-chromic etching: planar front along the FB replaced by cells after a few (< 10) micrometers, as well as solidification and migrated grain boundaries (dark lines).*

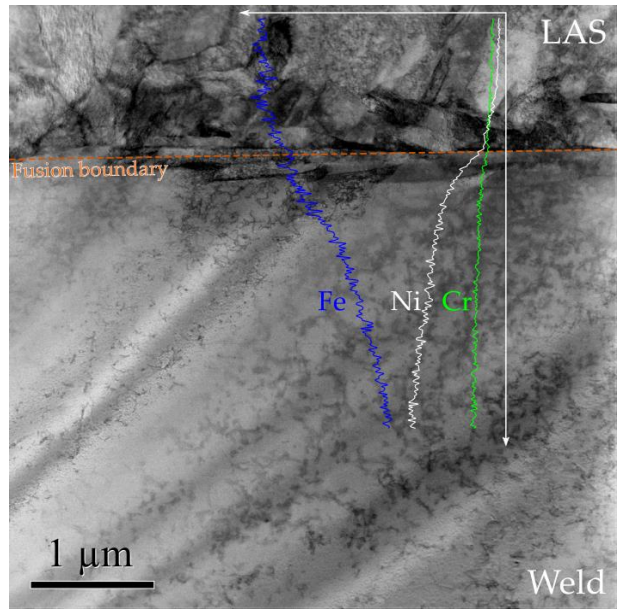

*Figure 5: Bright field STEM micrograph of the fusion boundary in the as-welded sample. The overlaid graph shows the evolution of the EDS signal associated to Fe, Ni and Cr. The evolution of the composition permits a clear identification of the position of the FB.*

1. As-welded

The microstructure of the weld right at the fusion boundary is shown in Figure 5. On the LAS side, the observed microstructure is consistent with fresh martensite during welding. The overlaid EDS signal permits to assess the shape of the composition gradient of the main substitutional elements, which in turn reveals the exact position the FB. The grains seem finer in a layer about 300 nm thick directly adjacent to the FB. The EDS signal reveals that there is a slight diffusion gradient on the LAS side, which extends over about the same distance. A thin layer of martensite can also be seen on the weld side, about 200 nm thick. Further, a single austenitic grain can be observed, containing numerous fine precipitates. Their appearance is cuboidal or parallelepipedal with a side length ranging from about 15 nm up to about 100 nm, as can be seen in the higher magnification micrograph in Figure 6a. From the associated EDS map in Figure 6b, it can be noted that those precipitates are Fe and Ni depleted while they are strongly enriched in Nb and Ti. EDS point analyses also revealed that they are enriched in Mo. Those three elements are strong carbide formers. The selected area electron diffraction (SAED) pattern given by the area in Figure 6a can be seen in Figure 7. The brightest reflections seen on the pattern are associated to the face centered cubic (FCC) matrix in [011] zone axis with a lattice parameter of about 3.6 Å. Much dimmer reflections can also be seen and correspond to an FCC structure with a lattice parameter of about 4.4 Å, both matching either NbC or TiC, in a cube-cube orientation relationship (OR) with the matrix. It can thus be assumed those precipitates are mixed niobium/titanium (Ti,Nb)C carbides , where part of the metallic atoms are substituted by molybdenum (Chen et al., 2014).

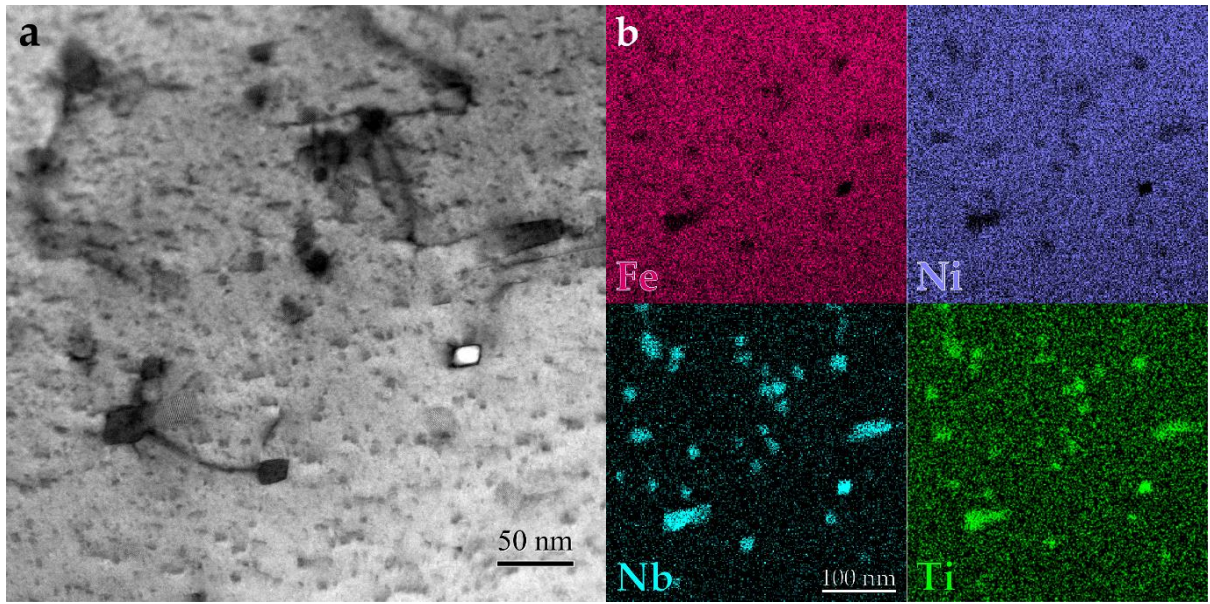

*Figure 6: a) Bright field STEM micrograph of the weld metal near the fusion boundary in the as-welded sample. b) EDS map of the corresponding area for Fe, Ni, Nb and Ti.*

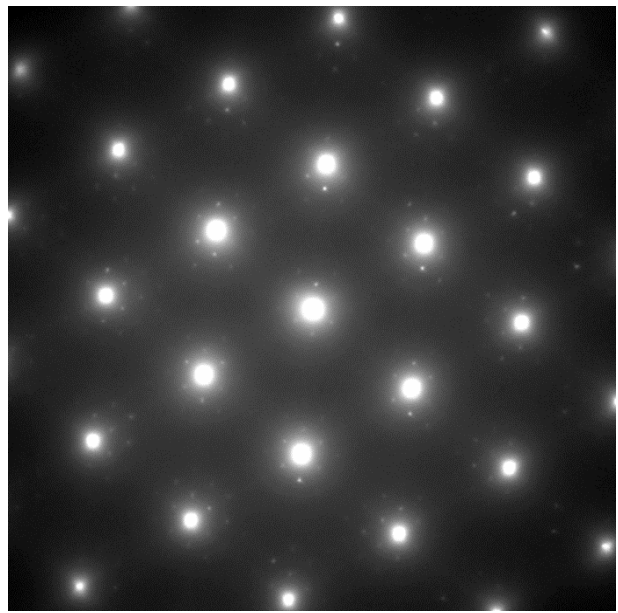

*Figure 7: Selected area electron diffraction pattern recorded from region shown in Figure 6. It shows the characteristic pattern of the FCC matrix in [011] zone axis. Dimmer reflections indicate the presence of another FCC phase in cube-cube orientation relationship, corresponding to (Ti,Nb)C. Reflections corresponding to matrix/precipitate double diffraction events are also visible.*

2. PWHT

The microstructure around the FB can be seen in Figure 8a, and two main changes can be noted. The first is the population of rather coarse precipitates appearing at the FB and at lath boundaries on the LAS side.

SEM observations suggest they are present in the as-welded state too but at a number density sufficiently lower that they are not found in Figure 5. As was previously observed (Parker and

Stratford, 2000), subsequent heat treatment, such as PWHT, leads to an increase of their number density, which made it more likely for them to be in the TEM specimen of Figure 6a.

The crystal structure of these precipitates was found to match $M_{23}C_6$ based on the diffraction pattern shown in and they were found to be Cr-enriched. All those precipitates were observed to be located at tempered martensite lath boundaries, including the ones at the FB due to the thin martensite layer present on the weld side that was also seen in the as-welded state.

A large fraction of nano-sized particles can also be seen precipitating intragranularly on the weld side, a few hundreds of nanometers away from the interface. The diffraction pattern in $[001]_\gamma$ zone axis seen in Figure 8b indicates the presence of an FCC phase in cube-cube OR with a lattice parameter about three times that of the matrix, corresponding to $M_{23}C_6$. It still shows reflections associated to (Ti,Nb)C as well.

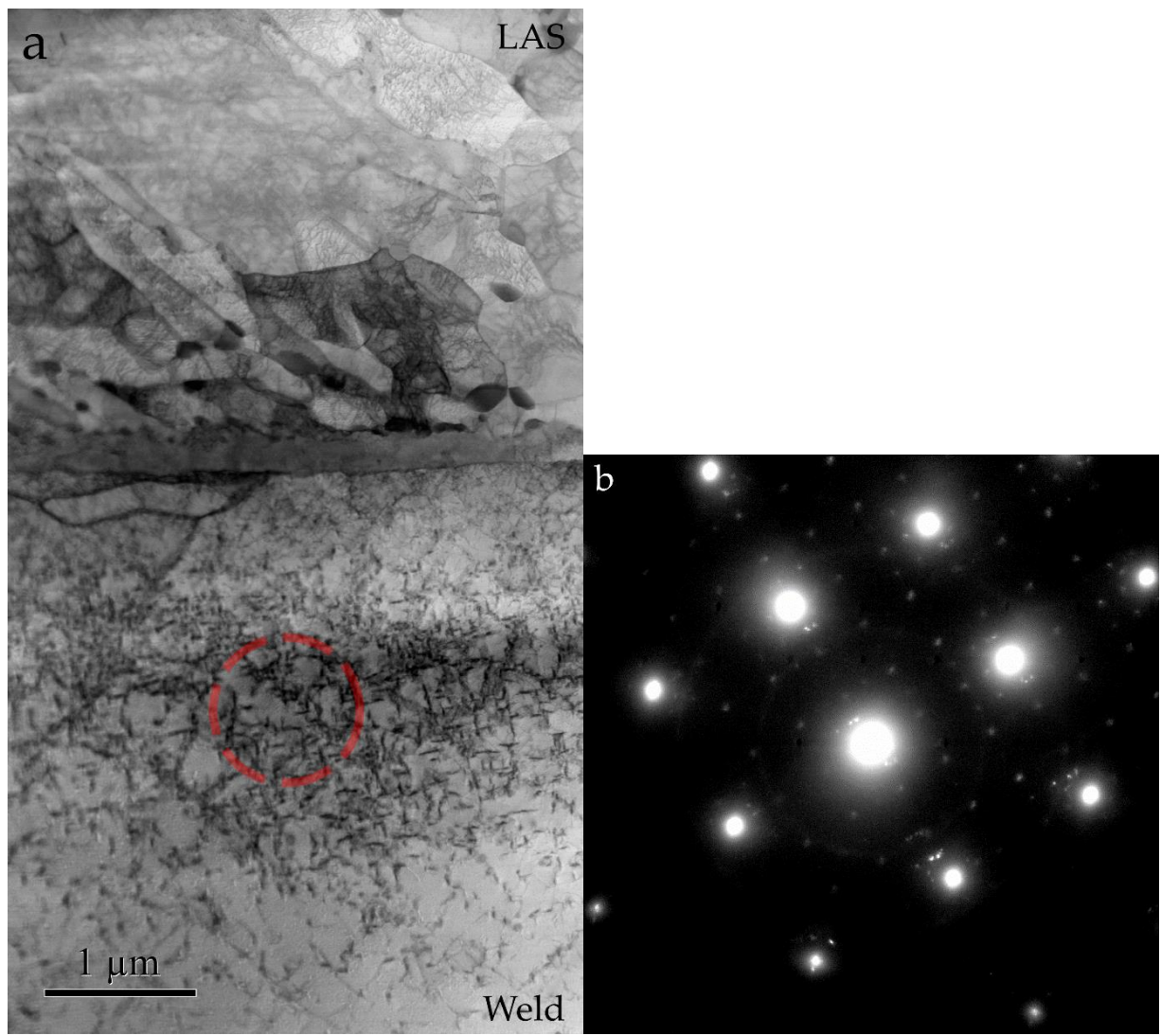

*Figure 8: a) BF STEM micrograph in PWHT sample 3.1. Carbides can be seen on lath boundaries on the LAS side and on the fusion boundary. A thin layer of martensite comparable to the as welded case can be seen on the weld side. However, numerous precipitates, previously absent, can be seen with their peak number density about 1 μm away from the fusion boundary. b) Selected area diffraction pattern from the precipiate rich area highlighted by the dashed red circle in a) on the weld side in $[001]_\gamma$ zone axis. It displays reflexions characteristic of both (Ti,Nb)C and $M_{23}C_6$, in addition to the ones of the matrix.*

## B. Composition profiles

The typical composition profiles for carbon and the main substitutional elements in the as-welded condition can be seen in Figure 9. Regarding substitutional elements, the composition on the steel side is constant and corresponds to the bulk content of 11CrMo9-10. Past the FB, there exists a composition gradient with increasing Cr, Mn and Ni content. This gradient zone is known as the partially mixed zone (PMZ) and is present in any DMW (DuPont, 2012). Further in the weld, those content values stabilize in the fully mixed zone (FMZ) but it should be noted that they never reach the bulk composition of alloy 82. Based on this stabilized composition far from the FB, a dilution coefficient can be calculated at about 4 % of LAS. This low dilution coefficient reveals that the metal filler feed rate was high with respect to the heat input of the process (DuPont and Marder, 1996). The size of the PMZ varies along the FB, bead to bead and within beads. However, the general shapes of the substitutional gradients are the same across all samples and locations, and are not affected in a detectable manner by the PWHT used here.

### 1. As-welded

Figure 10 shows a collection of carbon profiles recorded at different locations in as-welded samples M1 and M2. Those locations were separated in two groups showing either a large WL with a size between 17 µm and 27 µm or a small WL with a size between 9 µm and 17 µm. Those areas display PMZ measuring between 47 µm and 50 µm, and between 23 µm and 34 µm, respectively. In the as-welded condition, it can be seen that all carbon profiles are very similar. Up until 100 µm before the FB, the concentrations found on the LAS side correspond roughly to the nominal bulk composition. In the 100 µm before the FB, a decreasing gradient in carbon concentration indicates carbon transport from the body centered cubic (BCC) LAS side to the FCC weld side. In this same region, the profile looks very chaotic. This aspect can be attributed to the tempering of the LAS caused by the heat input from the welding process. This tempering leads to the formation of carbides, which in turn, cause large fluctuations of the local carbon concentration depending on whether or not they are caught by the EPMA probe. Right against the FB on the weld side, small carbon peaks can be distinguished in cases M1 P4 and M2 P2. Further in the weld, carbon concentration drops back down to zero with sudden and large fluctuations ascribed to the presence of coarse titanium/niobium carbides that were observed in the microstructure of the weld.

### 2. PWHT

In Figure 11, it can be seen that after PWHT the carbon profiles change little on the LAS side. The data in the decarburized zone seem less spread, suggesting that the carbides present after welding are dissolving. On the weld side, however, it can be noted that a carbon peak is systematically present within the first ten microns. Content values at the peak vary significantly from one location to another, from 0.25 wt.% C for M3.1 P3 and M3.2 P3 to over 0.6 wt.% C for M3.2 P5. Past the peak, carbon concentration decreases back to non-detectable contents between 30 and 40 µm away from the interface. The local large fluctuations owed to titanium/niobium carbides already seen in the as-welded samples are still present.

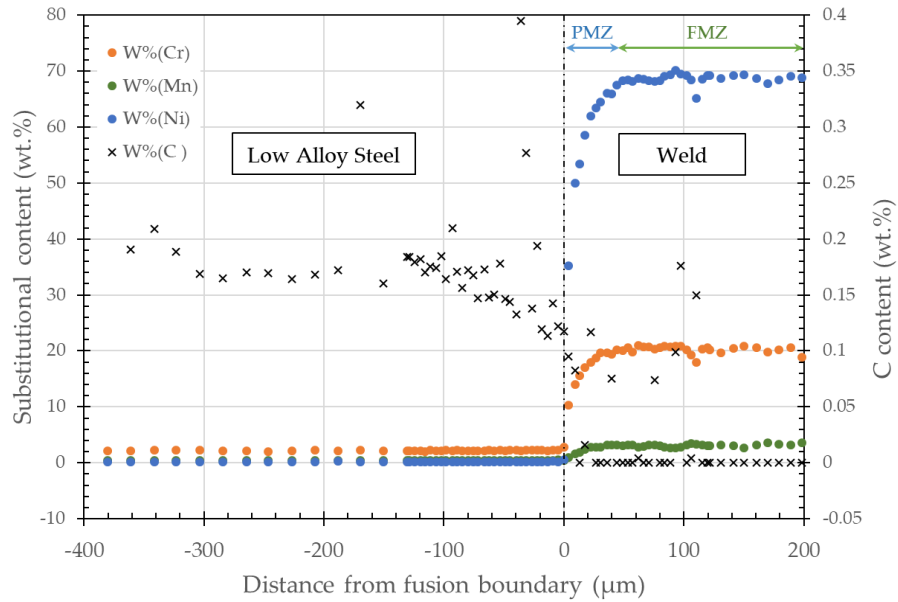

*Figure 9: Composition profile across the fusion boundary in sample 1 (as-welded) measured using EPMA. The overall shape of the substitutional profile is representative of the present dissimilar metal welds, although the length of the partially mixed zone varies along the fusion boundary. A slightly decarburized area, about 100 µm long, can be seen on the low-alloy steel side and is a typical feature as well.*

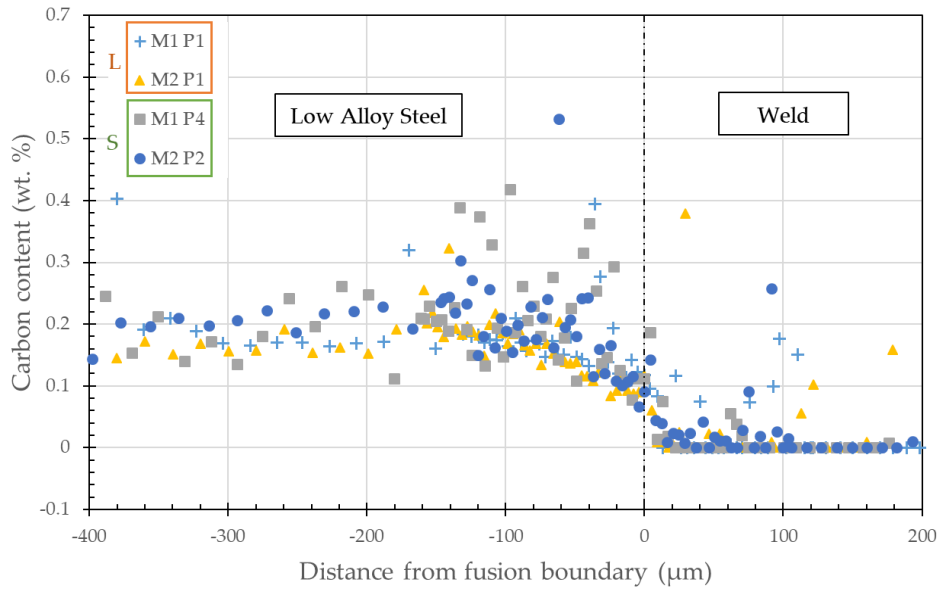

*Figure 10: Carbon composition profiles recorded across the fusion boundary at multiple locations in as-welded samples 1 and 2. The overall shapes of all profiles are very similar. Locations noted L (large) feature a white layer ranging from 17 µm to 27 µm in size, while those noted S (small) present a white layer between 9 µm and 17 µm wide.*

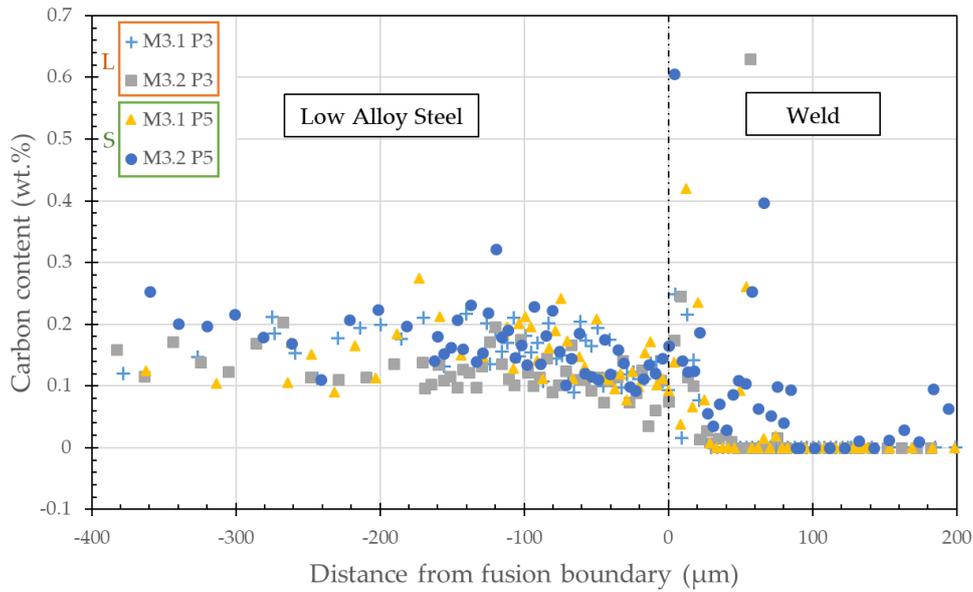

*Figure 11: Carbon composition profiles recorded across the fusion boundary at multiple locations in post-weld heat-treated samples 3.1 and 3.2. Locations noted L (large) feature a PMZ ranging from 17 µm to 27 µm in size, while those noted S (small) present a PMZ between 9 µm and 17 µm wide.*

### C. Mechanical properties

As covered in the previous sections, changes in the microstructure of the weld can occur over short length scale. In order to assess the evolution of the mechanical properties, nanoindentation hardness testing was employed to obtain measurements at coarse (20 µm matrix pitch over 600 µm) and fine spatial resolutions (2 µm matrix pitch over 40 µm) across the FB. The distance of individual indents from the FB is obtained by high-resolution optical imaging and the LAS is positioned on the left side (negative distance from the origin being the FB). Hardness values are reported at two different depths, 370 nm and 110 nm for coarse and fine matrices respectively. Since hardness was found to be depth-dependent, only relative changes between various locations and metallurgical states are commented.

#### 1. As-welded

Figure 12 shows the hardness profiles for the two as-welded samples in the identified ROIs. The measurements reveal important hardness variations depending on the position of the bead in the stack. Accordingly, the reference measurements far from the FB show that the LAS displays a pronounced mechanical softening, evolving from 6.2 GPa in average next to the top beads (M1 P1, M2 P1) down to 4 GPa for lower ones (M1 P4, M2 P2). The weld metal exhibits the opposite behavior and slightly hardens, from higher to lower beads.

Moreover, a distinctive hardness profile was obtained for higher beads in contrast with others: a sharp minimum of hardness is located at the interface, characterized by a steep decrease of hardness in the LAS part around 60 µm ahead the interface and an increase in the weld over 30 µm beyond the interface. Lower beads exhibit a much smoother profile, associating the interface with a shallower hardness minimum.

#### 2. PWHT

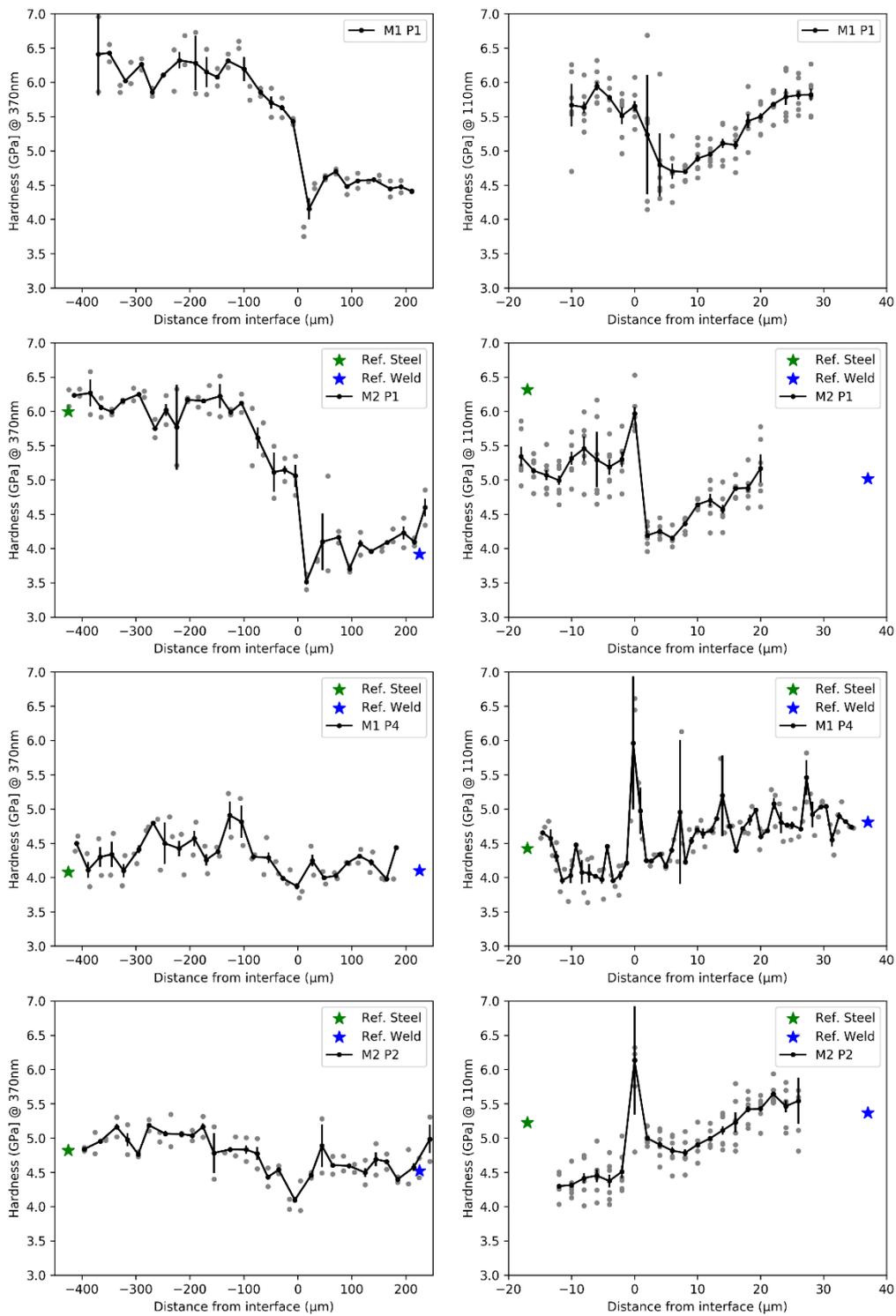

Figure 12: Nano-hardness profiles for the two as-welded samples M1 and M2. The left side column shows coarse matrix measurements (370 nm deep), while the right side one reports the fine matrix measurements (110 nm deep), closely arranged around the fusion boundary.

Figure 13 reports the hardness profiles of the two PWHT samples, M3.1 and M3.2. Reference matrices far from the FB reveal that the LAS component is softened to hardness values lower than the ones of the weld metal, which is observed to harden as compared to the as welded values. These profiles exhibit reproducible shapes. The LAS side exhibits an almost flat hardness profile. Past the FB, hardness rapidly increases to a maximum that is reached within 4 to 12 µm from the interface. Beyond it decreases over a short distance (20-30 µm) from this maximum value down to its reference level.

Finally, the data from the mechanical testing of all samples can be gathered on a so called 'phase map' (Randall et al., 2009): statistics of Hardness and Elastic modulus (H,E) measurements form clusters that uniquely identify a given composition associated with its microstructure. This representation is shown in Figure 14, reporting only the coarse matrix (H,E) measurements for clarity's sake. It can be seen that the LAS hardness increases with the depth in the stack of the neighboring bead in the as welded state, at almost constant elastic modulus. After PWHT, further softening is observed but the dependence on depth in stack is no longer statistically distinguishable. On the other hand, the weld side shows some hardening with increasing depth in stack along with a slight increase of elastic properties: this could be interpreted as the signature of phase transformation, like ordering or precipitation, taking place in the weld metal. The weld metal hardens and stiffens even more after PWHT, but bead-to-bead disparities seem to be smoothed out.

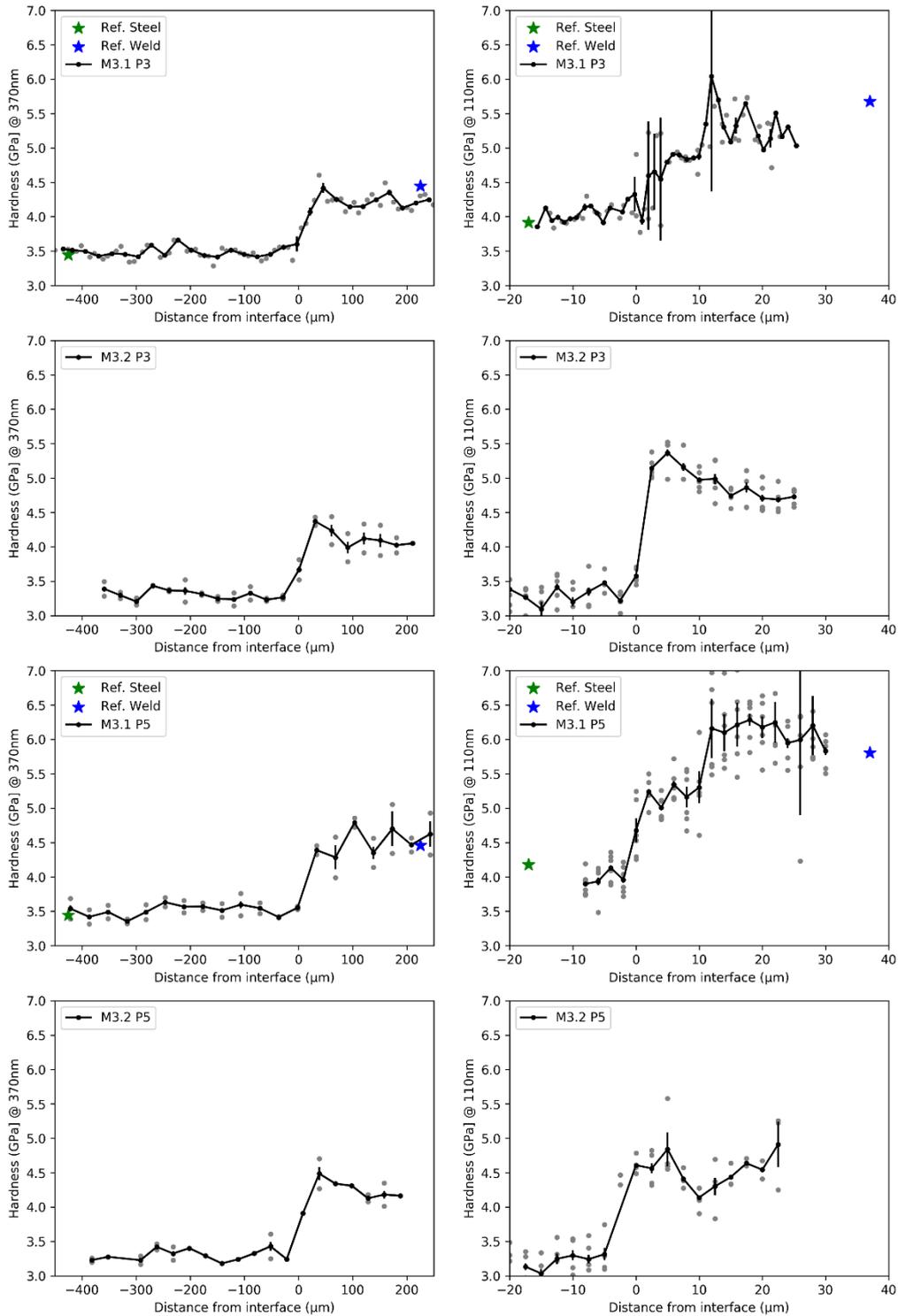

*Figure 13: Nano-hardness profiles for the two post-weld heat-treated samples M3.1 and M3.2. The left side column shows coarse matrix measurements (370 nm deep), while the right side one reports the fine matrix measurements (110 nm deep), closely arranged around the fusion boundary.*

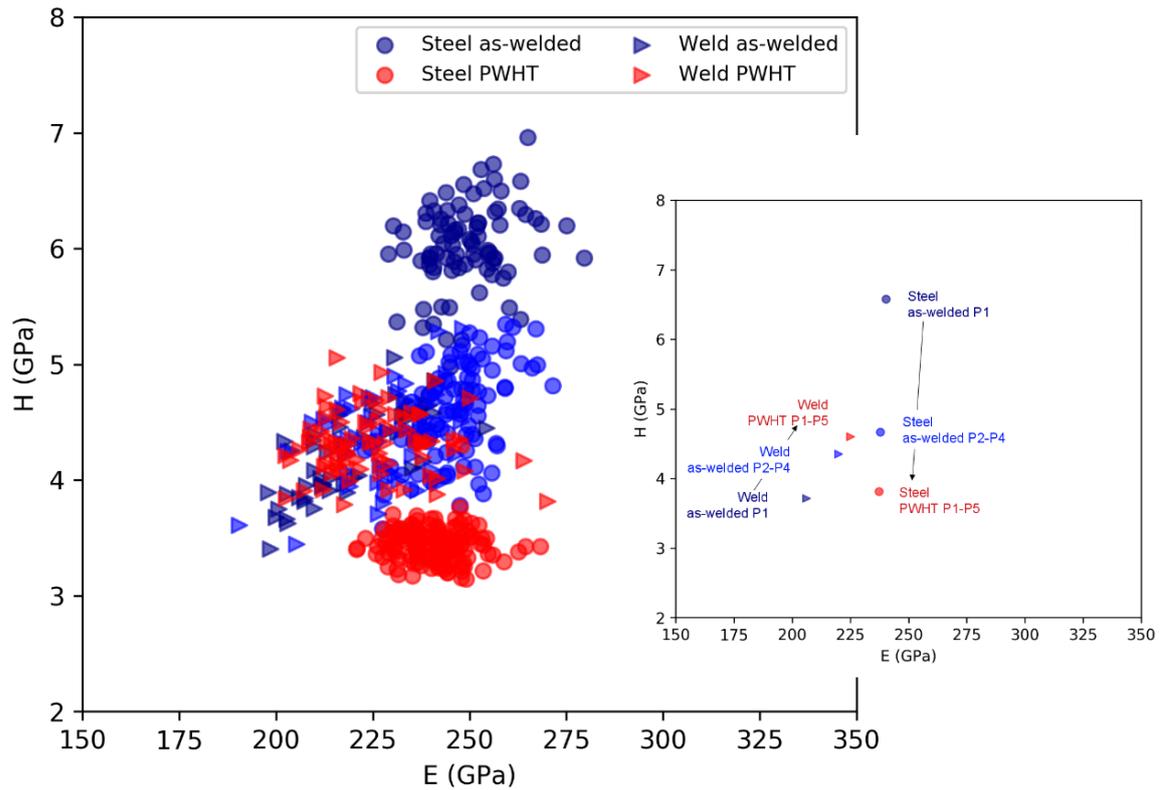

*Figure 14: Scatter plot of mechanical property measurements in all regions of interest of all samples (coarse matrices) in (Elastic modulus, Hardness) map. The inset details the statistical cluster evolution depending on heat treatment/annealing conditions.*

# IV. Discussion
## A. ROI selection

As has already been noted previously in the literature, DMWs produced using NG-GTAW present a very heterogeneous microstructure around the LAS/filler FB(Ahonen et al., 2018). The FB evolves from a rather straight continuous line at the top of a bead and becomes a feathery structure that becomes over 100 µm wide towards the bottom of the bead. The endeavor of the present paper is to study the effect of PWHT on the microstructure and the properties of the weld. In order to assess what this effect is, it is necessary to ascertain that the observed variations are owed to the PWHT and not simply to local fluctuations inherent to the welding process. To this end, the feathery regions will be disregarded here, notably because they make for a smaller portion of the FB. In addition, the comparison between our regions of interest was only considered receivable if the thicknesses of the WLs seen in optical micrographs were similar and if the shapes of the composition gradients of substitutional elements in the PMZ were comparable as well. The shape of the composition gradient in the PMZ was assessed using semi-quantitative EDS analysis in SEM in order to rapidly select the ROIs used for EPMA. In all cases, excellent agreement was found between SEM-EDS and EPMA results for substitutional elements.

## B. Importance of bead position in the stack

Even though NG-GTAW is generally considered a low-heat input technique(Ahonen et al., 2016), multiple indications have been found in this study to support an influence of the position of the bead in the stack on the surrounding microstructure and its associated properties. Part of this evidence can be found in the carbon profiles. First, in the as-welded condition, it can be seen that the apparent scatter of the data point in the decarburized region is maximal against the beads found at the low positions in both samples. Additionally, it is in those lowest positions that there seems to be a small carbon peak on the weld side when none can be seen in the upper positions. Accordingly, the nano-hardness profiles confirm a superior hardening on the weld side of the FB in beads that are located lower in the stack combined with greater softening in the decarburized region. These features are carried over to the PWHT state, although the hardness differences on the LAS side tend to level out. The influence of the position of the bead is also confirmed by micro-hardness measurements where it was found that the hardness of the LAS HAZ increases with the number of the adjoining bead in the as-welded state.

All these observations indicate that the extra thermal input from the overlying beads is sufficient to cause tempering differences on the LAS side and to induce differences in the progress of carbon diffusion. This implies that when comparing fine scale characterization results in different DMWs, the position of the considered bead in the stack matters as well, on top of the various criteria outlined in IV.A, in particular in the as-welded state.

### C. Carbon distribution across the fusion boundary

In the present DMW, the results shown in Figure 10 suggest that limited transport of carbon occurs during the welding process even when taking into account the elements in IV.B. Greater amounts of energy, such as the ones encountered during PWHT or in service, are typically required for the activation of carbon diffusion (Brentrup and Dupont, 2013).

In the as welded state, there exists a decarburized region as seen in Figure 10, even though no corresponding enrichment is seen on the weld side. This region of the LAS is heated to a temperature near the fusion point during welding. At this temperature, carbon transport will be extremely fast. The carbon from this region can diffuse to the FB where the higher chromium lowers its activity. Then, it can transfer to the liquid and distributes in the molten pool, where it would then react with Nb to precipitate as NbC. At the onset of solidification, it accumulates in the close vicinity of the FB and precipitates as $M_{23}C_6$. It could also accumulate in the thin layer of martensite on the weld side. In all cases, it would be concentrated in regions of the microstructure where its total amount would be underestimated in the EPMA profile due to the spatial resolution of the technique.

The presence of the micron-scale coarse NbC carbides is owed to the solidification mechanism. Such precipitates, several tens of micrometers in size, were characterized by SEM-EDS in the intercellular regions of the weld. They are also evidenced by the high local contents recorded on the weld side far from the FB, in the EPMA profiles in Figure 10 and Figure 11. Carbon, both from diffusion and dilution, and niobium accumulate in the liquid ahead of the solidification front leading to high local enrichment between cells where the last remaining liquid is thought to solidify into a mixture of those carbides and austenite.

Given their size and their homogeneous distribution, the cuboidal nano-scale (Ti,Nb)C carbides are thought to precipitate from solid state, when the beads are reheated by subsequent passes. The high Nb content of the filler metal results in an extremely low equilibrium carbon content in the austenitic matrix. For instance, taking a dilution coefficient of about 60 % LAS to represent the first 4 µm past the FB, the equilibrium carbon content is calculated to be $6 \times 10^{-4}$ wt.% at 900°C using the Thermo-Calc 2019a software package with the TCFE9 database.

In the case of ferritic/austenitic DMWs, there exists a driving force for carbon to diffuse from the ferritic LAS to the austenitic weld metal (Brentrup et al., 2012). It can be seen that the PWHT conditions chosen in the present work, 700°C for 2 h, lead to an appreciable transfer of carbon and to a simultaneous precipitation of carbides. The carbides that were found to be $M_{23}C_6$ are consistent with the rapidly increasing chromium content of the weld metal. Once it precipitates in this steep chromium gradient, the carbon is no longer mobile. Thus, carbon can only accumulate over the first 10-15 µm on the weld side, albeit up to substantial contents (0.6 wt.%) as seen in Figure 11.

### D. Effects on the mechanical properties

These changes in the microstructure translate into differences in the local mechanical properties.

In the as–welded state, it can be seen that there exists differences in the hardness profiles along the FB, as was the case for the carbon profiles. Hundreds of microns away from the FB, the martensite of the LAS HAZ presents a hardness between about 5.5 and 6.5 GPa. The large spread of those hardness values is thought to be related to the inhomogeneities along the FB inherent to the base metal, the process as well as to the position of the bead in the stack as discussed in IV.B. The hardness of the martensite drops over the last 100 µm before the FB, which is consistent with the observed decarburized zone.

The evolution of the hardness in the weld metal is rather consistent across all passes. The point at the interface can be harder due to the presence of carbide and chromium/carbon enriched martensite. The hardness starts from about 4.5 GPa where it is iron rich (high dilution coefficient) and becomes harder until it reaches the composition of the weld (low dilution coefficient) and a hardness of about 5.5 GPa.

In the as-welded state, it can be seen from the nano-hardness measurement that there exists a strong mechanical contrast between the martensite of the LAS HAZ and the PMZ of the weld, the former being harder than the latter. The soft zone seen in the first ≈ 25 µm of the weld can be attributed to the dependence of the austenite mechanical properties on the composition in substitutional elements. This kind of contrast is thought to promote failure in DMWs, where the crack deviates to the lower strength material(Nevasmaa et al., 2013). The configuration of the mechanical properties seen here is consistent with previous work where plastic deformation localization and crack propagation had been observed to occur in the weld metal in as-welded 2.25Cr-1Mo/alloy 82 NG-DMWs (Nivas et al., 2017), although it occurred further than the extent of the present soft zone. It is possible that this discrepancy is related to welding parameters, especially filler metal feed rate and heat input (DuPont, 2010).

In addition to the stress relief effect, PWHT leads to a tempering of the LAS and allows some of its carbon to transfer to the weld side where the resulting precipitation causes a strong local hardening. In the present PWHT conditions, tempering of this LAS grade is expected to be complete (Pilling and Ridley, 1982). The values are more even than the as welded condition, between about 3.5 and 4.0 GPa. Even though the decarburization was observed to be similar to the as-welded condition, its effect is not visible in the nano-hardness measurements. On the weld side, the carbon enrichment and its associated precipitation reactions are sufficient to level the hardness trough present in the as-welded state and even give rise to a hardness peak in the first ≈ 15 µm adjacent to the interface. The extent of those effects led here to an inversion of the aforementioned mechanical contrast. This inversed mechanical contrast can be just as detrimental to the fracture resistance of the weld as it has already been observed that specific PWHT could lead to no improvement of the mechanical properties over the as-welded state in a different system (Ahonen et al., 2016). It should be noted here that the PWHT used here is in the upper range of usual PWHT for the 2.25Cr-1Mo / alloy 82 combination. Thus, it can be postulated that the present PWHT conditions led to a degree of over-annealing of the DMWs. Samples treated at a lower temperature are currently under investigation to ascertain this hypothesis.

## V. Conclusion

Multiple characterization techniques were leveraged to establish the effect of a 2 h 700°C PWHT on the microstructure of the near FB region in a 2.25Cr-1Mo / alloy 82 DMW.

First, the information acquired at multiple scales highlights the inhomogeneity of the FB and the importance of careful identification of ROIs for metallurgical studies. It was notably found that the position of the bead in the stack can significantly affect local measures such as hardness or carbon composition profiles across the FB.

It was observed that the LAS is already decarburized over ≈ 120 µm preceding the FB in the as-welded condition, even though little to no excess carbon is seen on the weld side of the FB. This carbon is thought to transfer in the liquid during welding and distribute in the bulk of the weld, resulting in low overall elevation of the weld carbon content. Some carbon was also seen to precipitate as micron scale $M_{23}C_6$ carbides at the FB. Further on the weld side, two populations of NbC were found: a coarse micro-scale one and an abundant nano-scale one. As expected, the decarburized region is responsible for a drop in hardness of the HAZ martensite and the presence of the carbides at the FB causes a hardness peak. Interestingly, it was also found that there exists a hardness trough in the first microns of the PMZ on the weld side, leading to an unwanted mechanical contrast between the hard martensite and soft PMZ.

In addition to complete tempering of the LAS, the present PWHT also caused carbon to solid-state transfer to the weld side, accumulate over about 15 to 20 µm past the FB, and lead to an abundant precipitation of chromium-rich nano-scale $M_{23}C_6$. This additional carbon transfer is small compared to the amount that has already diffused in the liquid pool, as the decarburized region remained similar. However, it leads to a substantial hardening of the region, compensating the hardness trough seen in the as welded condition and even forming a hardness peak. Combined with the softer martensite, this results in a mechanical contrast, equal but inversed with respect to the as-welded condition. A PWHT at lower temperature might permit to optimize the microstructure to yield a more even hardness profile.

## VI. Acknowledgements

Part of this work was performed with the support of the Center of Excellence of Multifunctional Architectured Materials "CEMAM" No. AN-10-LABX-44-01 funded by the "Investment for the Future" Program operated by the National Research Agency (ANR).